\documentclass[twocolumn]{aastex631}
\usepackage{natbib}
\usepackage{multirow}
\usepackage[disable,colorinlistoftodos]{todonotes}
\graphicspath{{./}{figures/}}
\usepackage{threeparttable}

\begin{document}

\title{Extending the short gamma-ray burst population from sub-threshold triggers in Fermi/GBM and GECAM data and its implications}

\shorttitle{Sub-threshold Short Gamma-Ray Burst}
\shortauthors{Cai et. al}

\author{Ce Cai\textsuperscript{*}}
\email{caice@hebtu.edu.cn}
\affil{College of Physics and Hebei Key Laboratory of Photophysics Research and Application, 
\\Hebei Normal University, Shijiazhuang, Hebei 050024, China}
\affil{Shijiazhuang Key Laboratory of Astronomy and Space Science, Hebei Normal University, Shijiazhuang, Hebei 050024, China}

\author{Shao-Lin Xiong\textsuperscript{*}}
\email{xiongsl@ihep.ac.cn}
\affil{State Key Laboratory for Particle Astrophysics, Institute of High Energy Physics, Chinese Academy of Sciences, \\19B Yuquan Road, Beijing 100049, China}

\author{Yan-Qiu Zhang}
\affil{State Key Laboratory for Particle Astrophysics, Institute of High Energy Physics, Chinese Academy of Sciences, \\19B Yuquan Road, Beijing 100049, China}
\affil{University of Chinese Academy of Sciences, Chinese Academy of Sciences, Beijing 100049, China}

\author{Jin-Peng Zhang}
\affil{State Key Laboratory for Particle Astrophysics, Institute of High Energy Physics, Chinese Academy of Sciences, \\19B Yuquan Road, Beijing 100049, China}
\affil{University of Chinese Academy of Sciences, Chinese Academy of Sciences, Beijing 100049, China}

\author{Ping Wang}
\affil{State Key Laboratory for Particle Astrophysics, Institute of High Energy Physics, Chinese Academy of Sciences, \\19B Yuquan Road, Beijing 100049, China}

\author{Yao-Guang Zheng}
\affil{Department of Physics, Northeastern University, Shenyang 110819, China}
\affil{Key Laboratory for Particle Astrophysics, Institute of High Energy Physics, Chinese Academy of Sciences, \\19B Yuquan Road, Beijing 100049, China}

\author{Shi-Jie Zheng}
\affil{State Key Laboratory for Particle Astrophysics, Institute of High Energy Physics, Chinese Academy of Sciences, \\19B Yuquan Road, Beijing 100049, China}

\author{Shuo Xiao}
\affil{School of Physics and Electronic Science, Guizhou Normal University, Guiyang 550001, China}
\affil{Guizhou Provincial Key Laboratory of Radio Astronomy and Data Processing, Guizhou Normal University, \\Guiyang 550001, China}


\author{Hao-Xuan Guo}%
\affil{State Key Laboratory for Particle Astrophysics, Institute of High Energy Physics, Chinese Academy of Sciences, \\19B Yuquan Road, Beijing 100049, China}
\affil{Department of Nuclear Science and Technology, School of Energy and Power Engineering,\\ Xi'an Jiaotong University, Xi'an 710049, China}

\author{Jia-Cong Liu}%
\affil{State Key Laboratory for Particle Astrophysics, Institute of High Energy Physics, Chinese Academy of Sciences, \\19B Yuquan Road, Beijing 100049, China}
\affil{University of Chinese Academy of Sciences, Chinese Academy of Sciences, Beijing 100049, China}

\author{Yang-Zhao Ren}
\affiliation{State Key Laboratory for Particle Astrophysics, Institute of High Energy Physics, Chinese Academy of Sciences, \\19B Yuquan Road, Beijing 100049, China}
\affiliation{School of Physical Science and Technology, Southwest Jiaotong University, Chengdu 611756, Sichuan, China}

\author{Wen-Jun Tan}
\affil{State Key Laboratory for Particle Astrophysics, Institute of High Energy Physics, Chinese Academy of Sciences, \\19B Yuquan Road, Beijing 100049, China}
\affil{University of Chinese Academy of Sciences, Chinese Academy of Sciences, Beijing 100049, China}

\author{Chen-Wei Wang}%
\affil{State Key Laboratory for Particle Astrophysics, Institute of High Energy Physics, Chinese Academy of Sciences, \\19B Yuquan Road, Beijing 100049, China}
\affil{University of Chinese Academy of Sciences, Chinese Academy of Sciences, Beijing 100049, China}

\author{Yue Wang}
\affil{State Key Laboratory for Particle Astrophysics, Institute of High Energy Physics, Chinese Academy of Sciences, \\19B Yuquan Road, Beijing 100049, China}
\affil{University of Chinese Academy of Sciences, Chinese Academy of Sciences, Beijing 100049, China}

\author{Sheng-Lun Xie}
\affil{State Key Laboratory for Particle Astrophysics, Institute of High Energy Physics, Chinese Academy of Sciences, \\19B Yuquan Road, Beijing 100049, China}
\affil{Institute of Astrophysics, Central China Normal University, Wuhan 430079, China}

\author{Wang-Chen Xue}%
\affil{State Key Laboratory for Particle Astrophysics, Institute of High Energy Physics, Chinese Academy of Sciences, \\19B Yuquan Road, Beijing 100049, China}
\affil{University of Chinese Academy of Sciences, Chinese Academy of Sciences, Beijing 100049, China}

\author{Zheng-Hang Yu}
\affil{State Key Laboratory for Particle Astrophysics, Institute of High Energy Physics, Chinese Academy of Sciences, \\19B Yuquan Road, Beijing 100049, China}
\affil{University of Chinese Academy of Sciences, Chinese Academy of Sciences, Beijing 100049, China}

\author{Peng Zhang}
\affil{State Key Laboratory for Particle Astrophysics, Institute of High Energy Physics, Chinese Academy of Sciences, \\19B Yuquan Road, Beijing 100049, China}
\affil{College of Electronic and Information Engineering, Tongji University, Shanghai 201804, China}

\author{Wen-Long Zhang}
\affil{State Key Laboratory for Particle Astrophysics, Institute of High Energy Physics, Chinese Academy of Sciences, \\19B Yuquan Road, Beijing 100049, China}
\affil{School of Physics and Physical Engineering, Qufu Normal University, Qufu, Shandong 273165, China}

\author{Chao Zheng}%
\affil{State Key Laboratory for Particle Astrophysics, Institute of High Energy Physics, Chinese Academy of Sciences, \\19B Yuquan Road, Beijing 100049, China}
\affil{University of Chinese Academy of Sciences, Chinese Academy of Sciences, Beijing 100049, China}


\author{Jia-Wei Luo}
\affil{College of Physics and Hebei Key Laboratory of Photophysics Research and Application, 
\\Hebei Normal University, Shijiazhuang, Hebei 050024, China}
\affil{Shijiazhuang Key Laboratory of Astronomy and Space Science, Hebei Normal University, Shijiazhuang, Hebei 050024, China}

\author{Shuai Zhang}
\affil{College of Physics and Hebei Key Laboratory of Photophysics Research and Application, 
\\Hebei Normal University, Shijiazhuang, Hebei 050024, China}
\affil{Shijiazhuang Key Laboratory of Astronomy and Space Science, Hebei Normal University, Shijiazhuang, Hebei 050024, China}


\author{Li-Ming Song}
\affil{State Key Laboratory for Particle Astrophysics, Institute of High Energy Physics, Chinese Academy of Sciences, \\19B Yuquan Road, Beijing 100049, China}

\author{Shuang-Nan Zhang}
\affil{State Key Laboratory for Particle Astrophysics, Institute of High Energy Physics, Chinese Academy of Sciences, \\19B Yuquan Road, Beijing 100049, China}
\affil{University of Chinese Academy of Sciences, Chinese Academy of Sciences, Beijing 100049, China}

\begin{abstract}
Detection of short gamma-ray bursts (SGRBs) is critically important for the research of compact object mergers and multi-messenger astrophysics, but a significant part of SGRBs fall below the trigger threshold of GRB detectors, and thus are often missed. Here we present a systematic search for and verification of missed SGRBs using Fermi/GBM subthreshold triggers, jointly analyzing data from GBM, GECAM-B, and GECAM-C.
Among 466 Fermi/GBM sub-threshold events (with reliability $\geq$5) from 2021 to 2024, 181 are within GECAM’s field of view. We find that 49 out of 181 are confirmed astrophysical transients, and 41 can be classified as SGRBs.
Thus, the SGRB detection rate of Fermi/GBM is increased to about 50 per year. Additionally, a complete multi-instrument monitoring and systematic verification of GBM sub-threshold events is expected to further increase the SGRB rate to about 80 per year, which is $\sim$100\% improvement relative to the GBM-triggered SGRBs. 
These results may have important implications on the local formation rate of SGRBs and the binary neutron star merger rate. 
We also searched for potential temporal coincidences between these SGRBs and gravitational waves
from the LIGO--Virgo--KAGRA O4 run resulting in no detection.
\end{abstract}


\section{Introduction}

The joint detection of GW170817 and GRB 170817A \citep{2017PhRvL.119p1101A, 2017ApJ...848L..14G, 2017ApJ...848L..15S} provided the first direct evidence linking binary neutron star (BNS) mergers to short gamma-ray bursts (SGRBs), and heralded the era of multi-messenger gravitational-wave astronomy. Electromagnetic (EM) counterparts such as GRBs offer critical insights into the physical processes underlying compact object mergers (e.g., \citealt{2017ApJ...848L..13A, 2017Sci...358.1559K, 2018Natur.561..355M}), highlighting the importance of gamma-ray observations in multi-messenger studies.

Importantly, GRB 170817A is among the weakest short GRBs that GBM has triggered on in terms of its 64 ms peak flux and is 2 to 6 orders of magnitude less energetic than other bursts with measured redshifts \citep{2017ApJ...848L..13A,2017ApJ...848L..14G}. It demonstrates the need for systematic searches to recover similar events at greater distances \citep{2018NatCo...9..447Z}.

Among current wide-field gamma-ray instruments, the Fermi Gamma-ray Burst Monitor (GBM; \citealt{2009ApJ...702..791M}) and the Gravitational Wave High-energy Electromagnetic Counterpart All-sky Monitor (GECAM; \citealt{li2022technology, ZHANG2023168586, 2024SCPMA..6711013F}) are among the most prolific all-sky monitors. Both of them have detected many SGRBs. For example, GBM triggers on approximately 40 SGRBs per year
\citep{2014ApJS..211...12G,2014ApJS..211...13V,2016ApJS..223...28N,2020ApJ...893...46V}. However, many weak or unfavorably oriented events fall below on-board trigger thresholds \citep{2021arXiv211205101Z}. To recover these sub-threshold signals, both missions have developed offline search pipelines that enhance sensitivity and enable deeper exploration of the transient gamma-ray sky \citep{2015ApJS..217....8B, 2018ApJ...862..152K, 2025SCPMA..6839511C}. 

However, low-significance events such as a candidate counterpart to the first direct observation of a binary black hole coalescence event, GW150914 \citep{2016PhRvD..93l2003A, 2016ApJ...826L...6C}, remain controversial in terms of its astrophysical origin \citep{2016ApJ...820L..36S, 2016ApJ...827L..38G, 2016arXiv160505447X}, underscoring the importance of coordinated multi-instrument analyses to improve sensitivity and detection confidence for sub-threshold events.

The Energetic Transients joint analysis system for Multi-INstrument (ETJASMIN) pipeline \citep{2022MNRAS.514.2397X, 2025ApJS..277....9C} enables coherent, likelihood-based searches across gamma-ray observatories, including GECAM-B, GECAM-C, and Fermi/GBM. Building on prior GBM and GECAM methodologies \citep{2015ApJS..217....8B, 2025SCPMA..6839511C}, it evaluates spatial and spectral consistency across up to 49 detectors to improve sensitivity, reduce false positives, and recover weak signals \citep{2025ApJS..277....9C}. Simulations indicate enhanced detection significance 
and source amplitude estimation relative to single-instrument searches. Applied recently to 63 X-ray transients from EP/WXT, the ETJASMIN pipeline identified gamma-ray counterparts in 22\% of X-ray transients and provided stringent upper limits for other X-ray transients, demonstrating its efficacy in characterizing the high energy emission property of soft X-ray transients \citep{2025arXiv250605920Z}.  

While substantial progress has been made in the detection of sub-threshold GRBs,
most studies have been restricted to single-instrument analyses, with limited cross-validation using independent observatories. The Fermi/GBM has published a catalog of sub-threshold triggers with uncertain astrophysical origin\footnote{\url{https://gcn.gsfc.nasa.gov/gcn/fermi_gbm_subthresh_archive.html}}. No systematic multi-instrument study has assessed how many of these candidates represent real astrophysical transients or even SGRB.

Therefore, in this work we apply the ETJASMIN pipeline to analyze sub-threshold candidates from the GBM sub-threshold trigger catalog. We systematically examine whether weak transient signals in GBM data can be found and verified through temporal and spatial coincidence with GECAM observations. 

This paper is organized as follows. Section \ref{sec:instrument} provides an overview of the GBM and GECAM instruments. Section \ref{sec:sample} describes the sample selection, and Section \ref{sec:analysis} details the data analysis methods. Section \ref{sec:results} presents the main results,  followed by the discussion and conclusions in Section \ref{sec:discussion} and \ref{sec:summary}.

\section{Instruments}
\label{sec:instrument}

\subsection{Fermi Gamma-ray Burst Monitor}
\label{sec:gbm}
The Gamma-ray Burst Monitor (GBM) onboard the Fermi Gamma-ray Space Telescope is a dedicated instrument for monitoring the gamma-ray sky in the energy range of approximately 8 keV to 40 MeV \citep{2009ApJ...702..791M}. It consists of 12 semidirectional sodium iodide (NaI) scintillation detectors, sensitive to photons between 8 and 1000 keV, and 2 bismuth germanate (BGO) detectors, covering a higher energy band from 200 keV to 40 MeV. The NaI detectors are oriented to view nearly the entire sky unocculted by Earth, while the BGO detectors are mounted on opposite sides of the spacecraft to ensure broad angular coverage.

\subsection{GECAM}
\label{sec:gecam} 
The Gravitational wave high-energy Electromagnetic Counterpart All-sky Monitor (GECAM) is a satellite constellation mission aimed at monitoring diverse types of high-energy transient events (e.g., \citealp{2022ApJ...935...10C,2023arXiv230301203A,2024SCPMA..6789511Z,Afterglow_zhengchao,GRB230307A_sunhui,mini_jet_yishuxu,Minimum_Variation_Timescales_xiao,TGF_TEB_zhaoyi}) across the sky in the energy range of $\sim$ 10 keV to 6 MeV.

The GECAM constellation currently comprises four satellites. The first pair, GECAM-A and GECAM-B, were launched into low Earth orbit (LEO) on 2020 December 10 \citep{li2022technology}. GECAM-C, also referred to as the High Energy Burst Researcher, was deployed on 2022 July 27 aboard the SATech-01 satellite \citep{2023NIMPA105668586Z}, and GECAM-D (also known as GTM) onboard the DRO-A satellite was launched into a distant retrograde orbit (DRO) on 2024 March 13 \citep{2024SCPMA..6711013F,2024ExA....57...26W}.
Each satellite carries gamma-ray detectors (GRDs; \citealt{an2022design}) for high-energy photon detection, and GECAM-A, -B, and -C are also equipped with charged particle detectors (CPDs; \citealt{xu2022design}), which aid in distinguishing between photon-driven bursts and charged particle events.
 Due to operational constraints, only GECAM-B and GECAM-C data are used for this study.

\section{Sample}
\label{sec:sample}

The Fermi/GBM sub-threshold trigger archive\footnote{\url{https://gcn.gsfc.nasa.gov/fermi\_gbm\_subthresh\_archive.html}} catalogs candidate transient events that failed to meet onboard trigger criteria but were subsequently identified through offline ground-based searches. Each entry provides the event time, sky localization, duration (i.e., the binning timescale that maximizes SNR), spectral hardness, and a reliability score (dubbed as REL) ranging from 1 to 10, indicating the likelihood of an astrophysical origin. 
At the time of analysis, the archive has accumulated 5614 events since its inception on 2017 April 16, including 3557 with REL $=$ 2 (63.4\%), 1348 with REL $=$ 5 (24.0\%), and 709 with REL $=$ 8 (12.6\%).

Among events with REL $=$ 5, 997 are classified as short bursts (duration $\leq 2$ s), corresponding to an average of $\sim$124 short sub-threshold candidates per year. For REL $=$ 8, 529 short bursts are identified, yielding an average rate of $\sim$66 per year over the same eight-year period.

To identify potential joint detections with GECAM, we compiled a sample of GBM sub-threshold events reported between 2021 February 1 and 2024 December 31, corresponding to GECAM’s operational period. We selected all events with REL $\geq 5$, yielding 466 candidates, including 227 with REL $=$ 8. We then applied visibility criteria: events were retained only if they occurred outside the South Atlantic Anomaly (SAA), during GECAM’s active observation periods, and were not Earth-occulted (based on the GBM-reported central sky position). We note that this approximation neglects localization uncertainty, because the complication on the systematic error of GBM burst prevents us from an accurate treatment.
This filtering resulted in a final sample of 181 events, including 102 with REL $=$ 8. For each event, we extracted the trigger time, duration, sky location, and REL from the GBM archive for further analysis.

\begin{figure}[htbp]
\epsscale{1}
\plotone{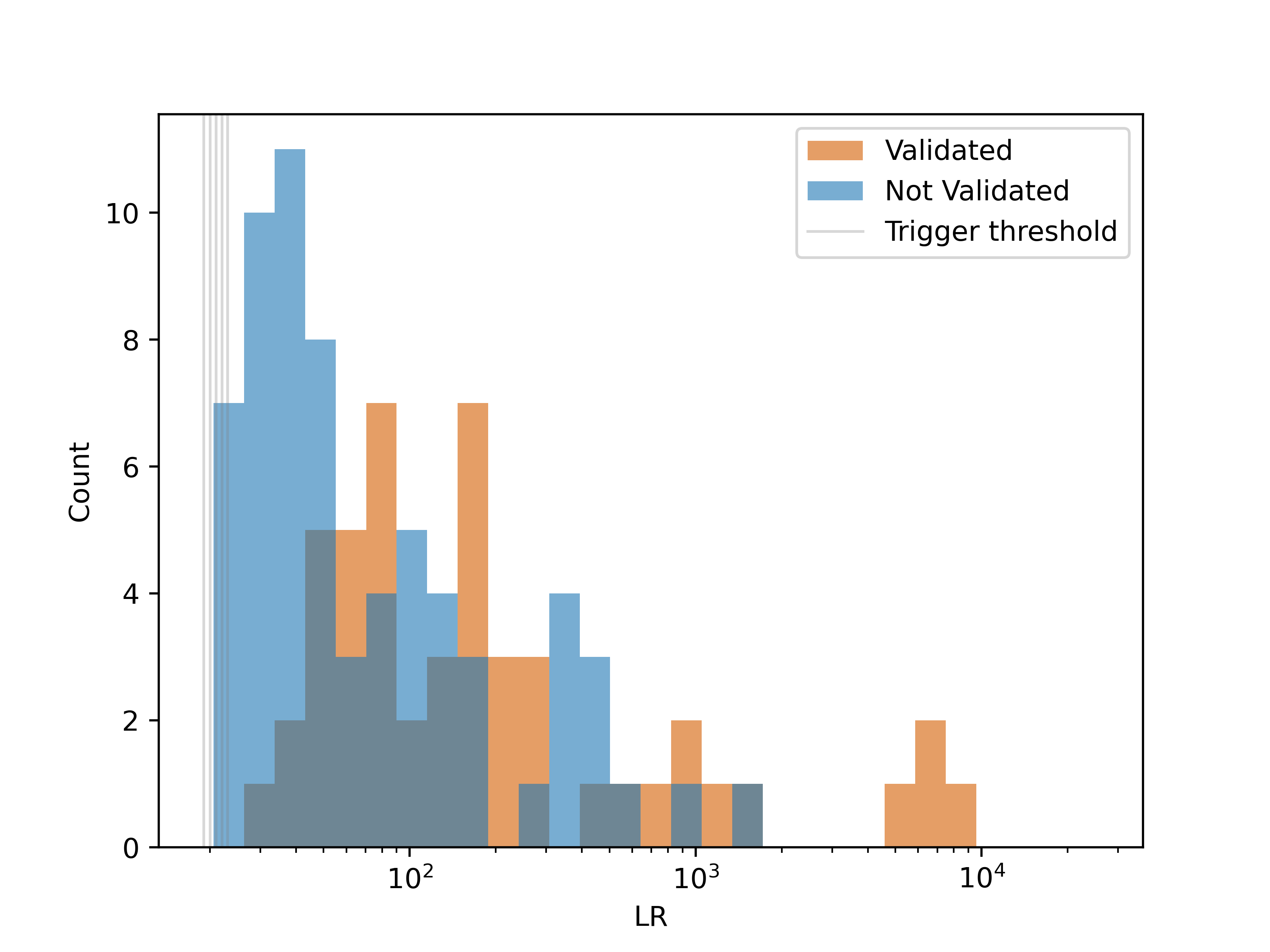}
\caption{Distribution of the joint likelihood ratio (LR) values for the 116 candidates exceeding the LR thresholds (marked in gray). The blue bars represent all events exceeding the predefined LR thresholds that were not confirmed by manual inspection, while the orange bars indicate those further confirmed through visual inspection.}
\label{fig10}
\end{figure}

\begin{figure}[htbp]
\epsscale{1}
\plotone{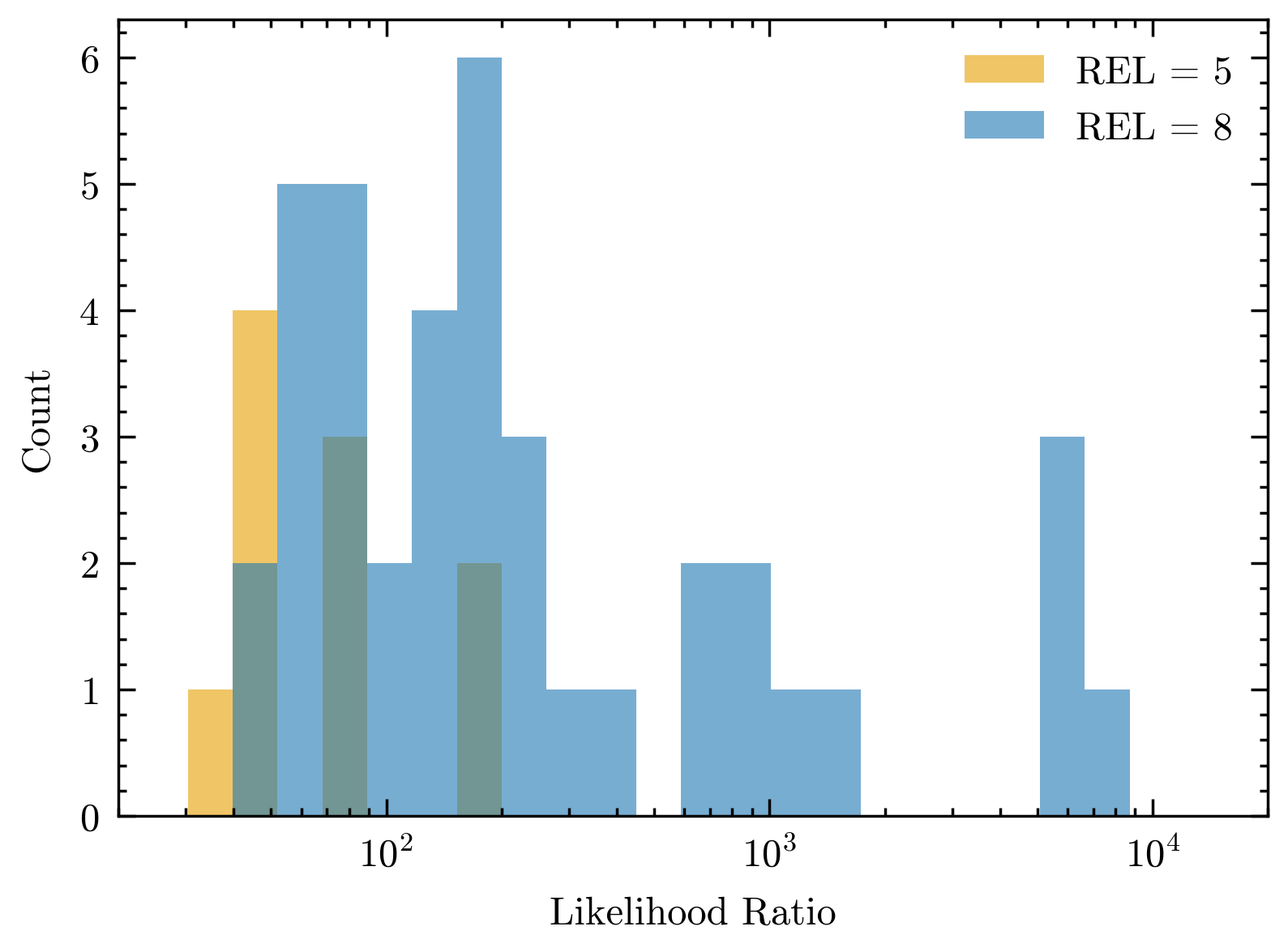}
\caption{Distribution of the likelihood ratio (LR) for the 49 validated events, separated by GBM reliability score. Events with a reliability score of 5 are shown in yellow, while those with a score of 8 are shown in blue.}
\label{fig1}
\end{figure}

\begin{figure}[htbp]
\epsscale{1}
\plotone{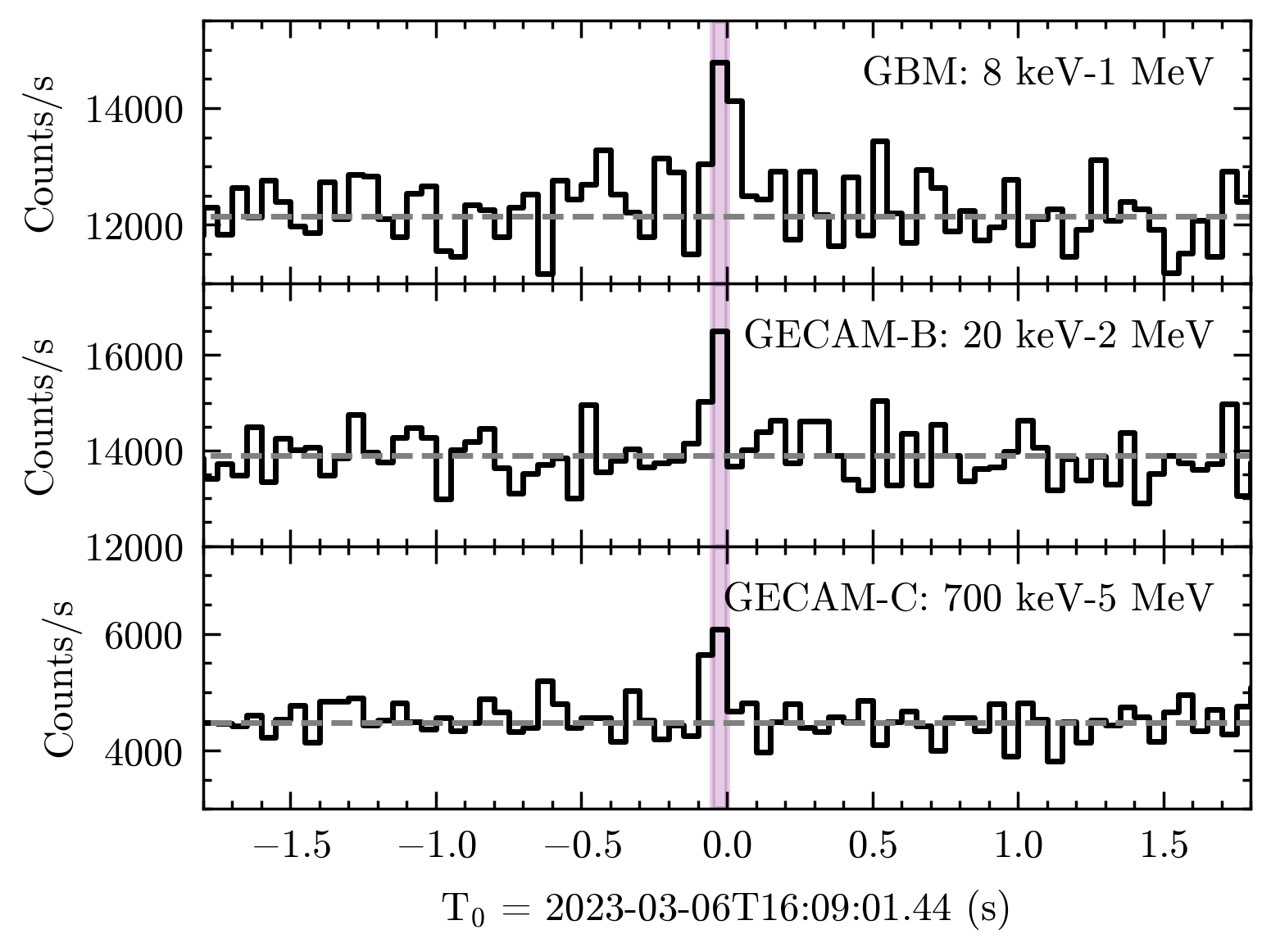}
\caption{Light curves of a representative burst detected jointly by GBM, GECAM-B, and GECAM-C. The event occurred at $T_0 = \text{2023-03-06T16:09:01.44}$ (UTC). Count rates are shown in 50 ms bins. The shaded region indicates the 
time bin that maximizes the joint signal significance (approximately 12 $\sigma$).}
\label{fig2}
\end{figure}

\begin{figure}[htbp]
\epsscale{1}
\plotone{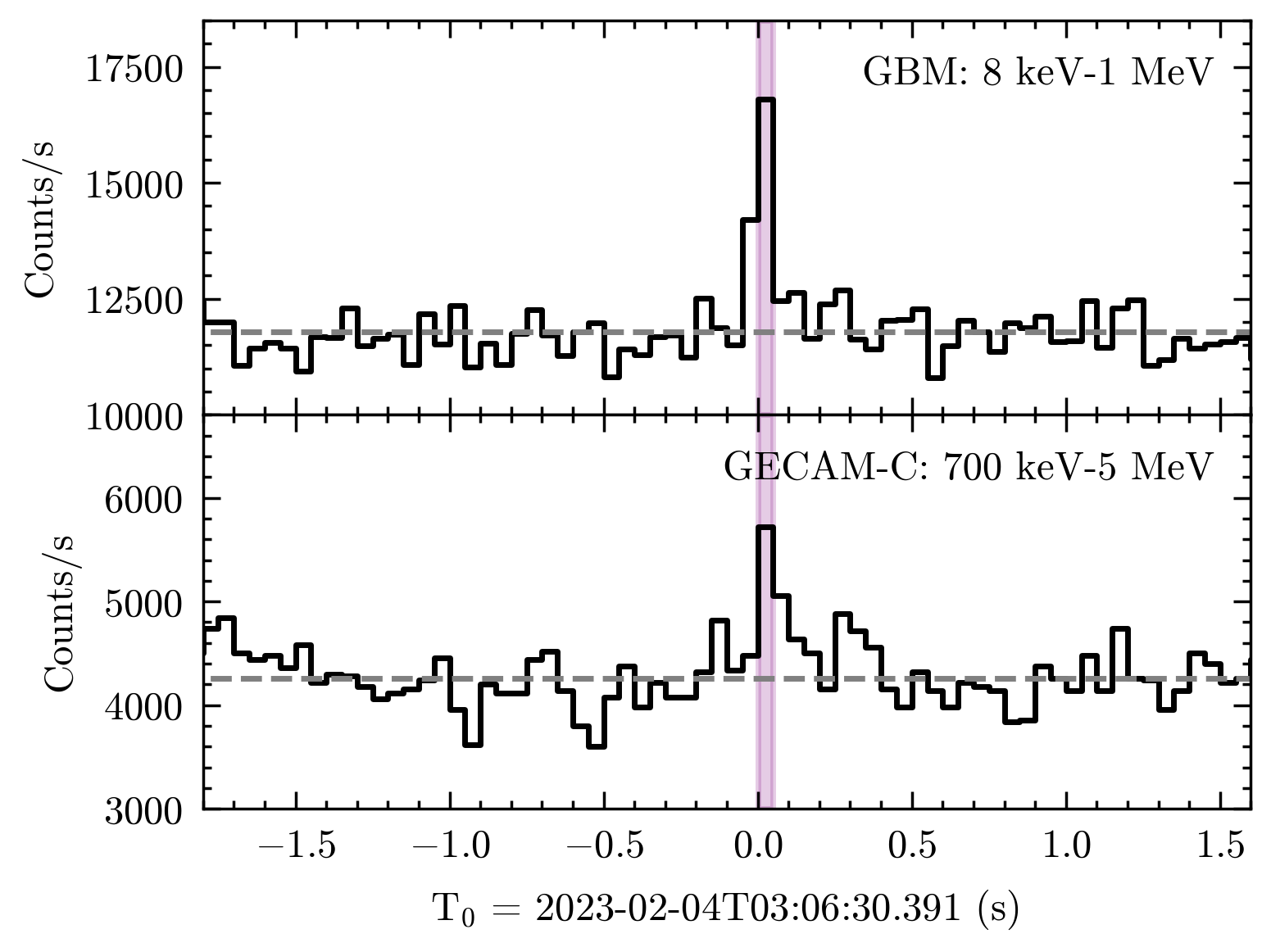}
\caption{Light curves of a sub-threshold burst detected jointly by GBM and GECAM-C. The event occurred at $T_0 = \text{2023-02-04T03:06:30.391}$ (UTC). Count rates are shown with 50 ms bins. The shaded region indicates the 
time bin that maximizes the joint signal significance (approximately 17 $\sigma$).}
\label{fig3}
\end{figure}

\begin{figure}[htbp]
\epsscale{1}
\plotone{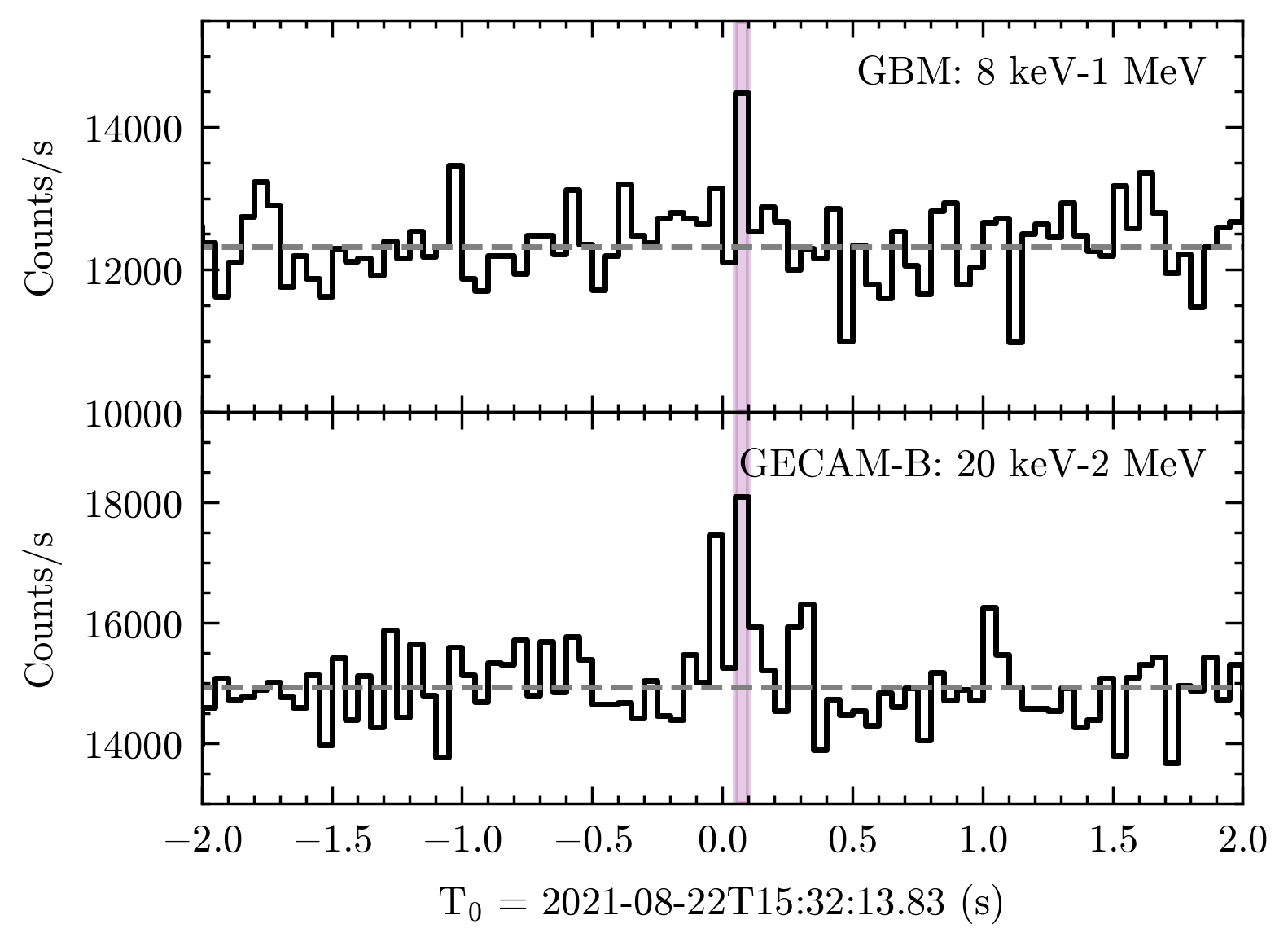}
\caption{Light curves of a sub-threshold burst detected by GBM and GECAM-B. The event occurred at $T_0 = \text{2021-08-22T15:32:13.83}$ (UTC). Count rates are plotted with 50 ms resolution. The shaded region indicates the 
time bin that maximizes the joint signal significance (approximately 12 $\sigma$).}
\label{fig4}
\end{figure}

\begin{figure*}[htbp]
\epsscale{1}
\plottwo{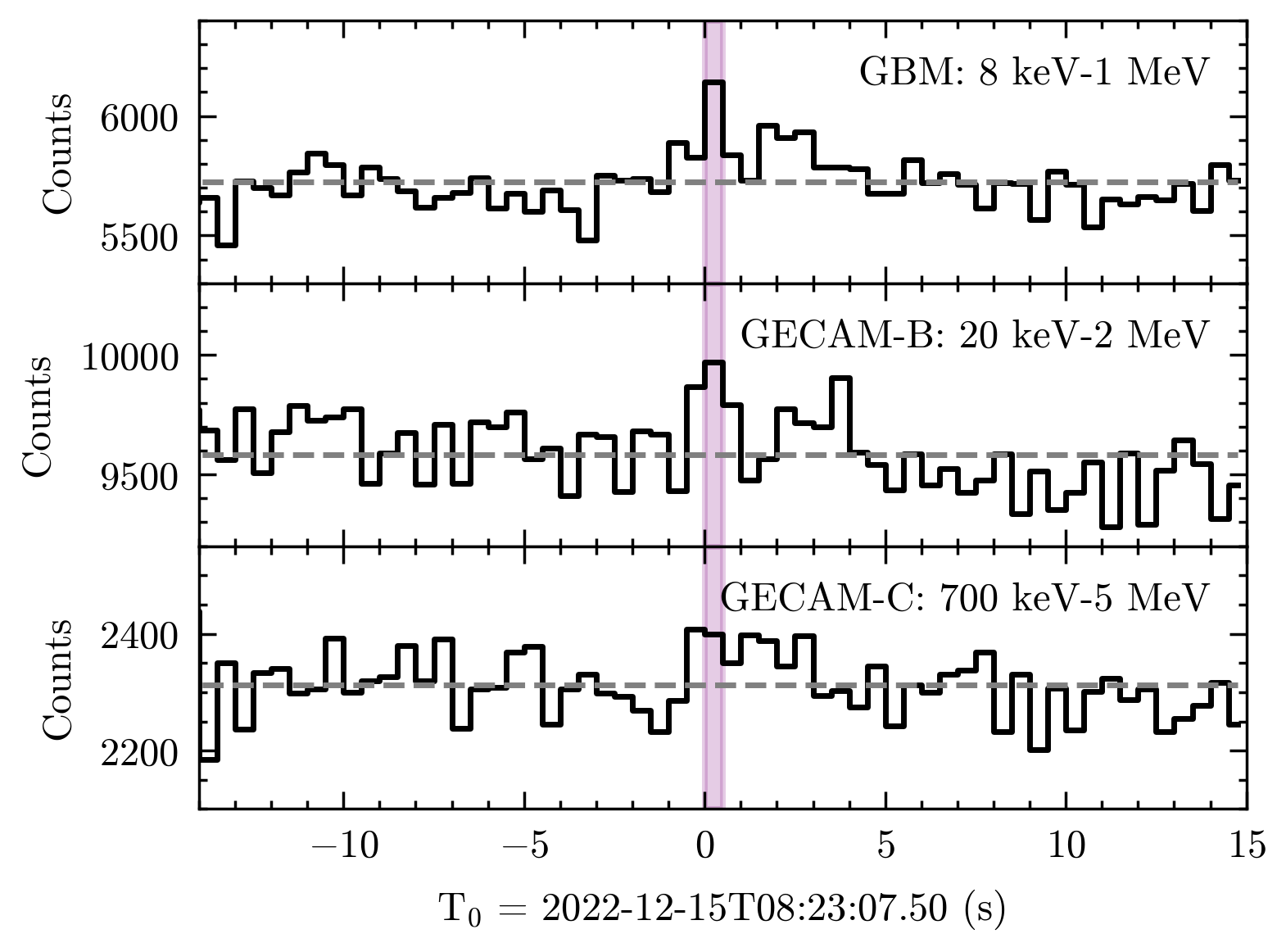}{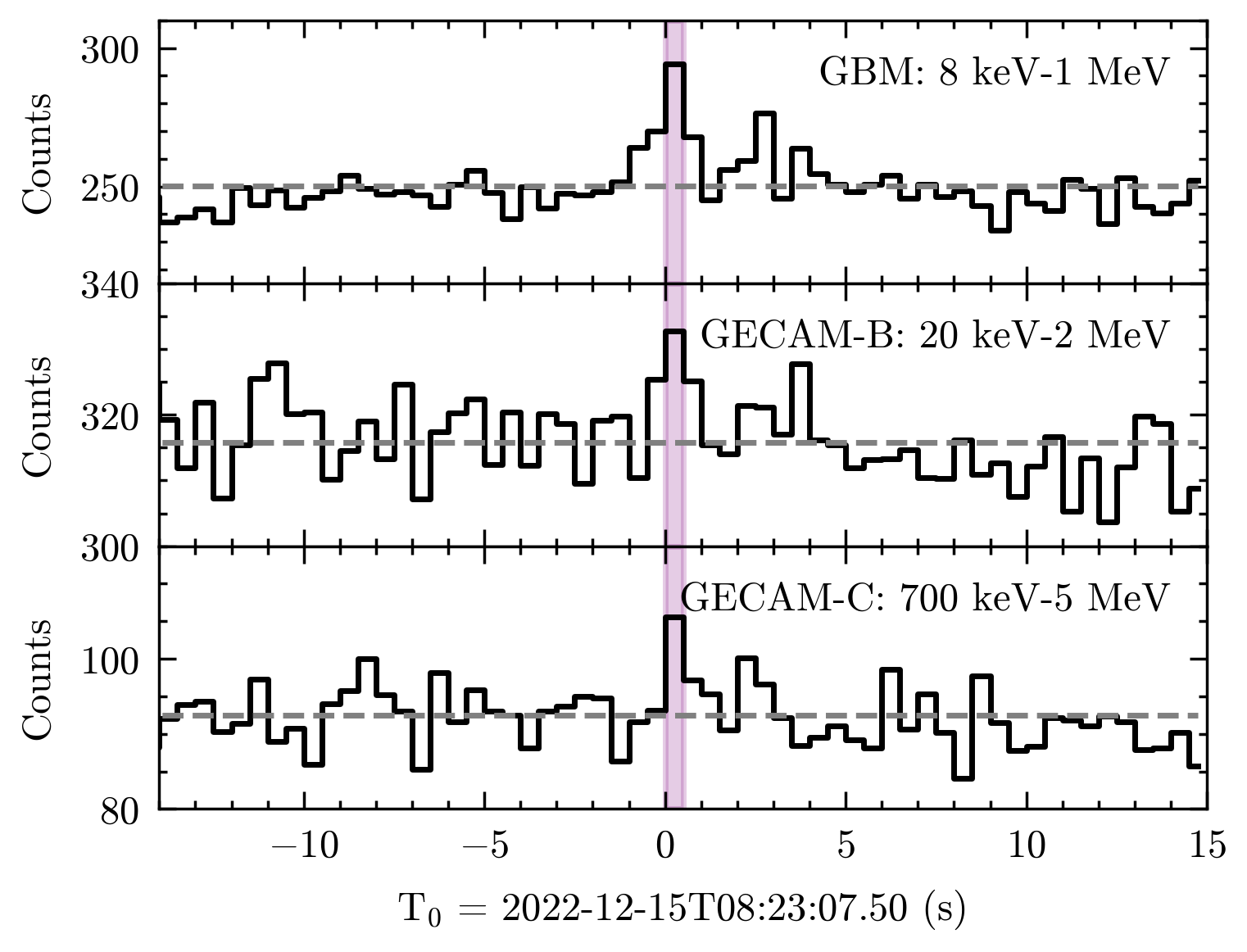}
\caption{Joint light curves of a sub-threshold burst detected on 2022 December 15 at $T_0 = \text{08:23:07.50}$ (UTC). The left panels show the summed counts from all detectors in GBM , GECAM-B, and GECAM-C. The right panels present the optimized light curves, where counts from all detectors are summed with different weight factors. The shaded region indicates the 
time bin that maximizes the joint signal significance (approximately 10 $\sigma$).}
\label{fig5}
\end{figure*}

\section{Analysis}
\label{sec:analysis}

\subsection{Burst Search}
To perform a targeted search for temporally coincident gamma-ray signals associated with GBM sub-threshold triggers, we employed the ETJASMIN targeted search pipeline \citep{2025ApJS..277....9C}. For each event in our sample, we analyzed a $\pm5$ s time window centered on the GBM trigger time using Time-Tagged Event (TTE) data from GBM and Event-Level (EVT) data from GECAM-B and GECAM-C.

The ETJASMIN framework integrates data from a total of 49 detectors, including 12 NaI detectors onboard GBM, 25 GRDs on GECAM-B, and 12 GRDs on GECAM-C. When available, data from all detectors are analyzed jointly to maximize sensitivity. This joint analysis is enabled by the similar detector responses among the NaI and GRD instruments, as demonstrated by in-flight cross-calibration between GECAM-B/C and Fermi/GBM \citep{2023arXiv230300698Z,2024RAA....24j4005Q}. It is further informed by satellite orbit and attitude parameters, which determine the source direction relative to each detector, and by precomputed instrument response functions.

The energy selection used in the analysis is tailored to each instrument. For GBM, we adopt the 8–1000 keV range. For GECAM-B and GECAM-C, the energy range is determined by the gain and bias voltage settings of each GRD \citep{ZHANG2022166222}, and calibrated response \citep{2023arXiv230300698Z,zhang2025grd} are used to select the optimal range separately for high-gain and low-gain modes.

The joint analysis compares the observed count rates to expectations under two competing hypotheses: signal-plus-background versus background-only. Specifically, the probability for each detector $k$ to measure the observed data under the signal-plus-background hypothesis (${\rm H}_{1}$) is given by:
\begin{equation}
{{P}_{k}({d}_{k}|{\rm H}_{1})}= \prod_{i} \frac{1}{\sqrt{2\pi}{\sigma}_{{d}_{i}}}exp(-\frac{({\widetilde{d}_{i}-{r}_{i}s)^2}}{2\sigma^{2}_{{d}_{i}}}),
\end{equation}
where the product is carried out over each channel $i$, ${n}_{i}$ represents the estimated background, $\widetilde{d}_{i}={d}_{i}-\left\langle{n}_{i}\right\rangle$ is the background-subtracted counts, ${r}_{i}$ represents the expected counts with a default amplitude of 1, obtained by multiplying the source models\footnote{The source models are Band functions \citep{1993ApJ...413..281B} with soft, normal, and hard parameter sets adopted from \citet{2015ApJS..216...32C}, corresponding to the templates RspSoft, RspNorm, and RspHard listed in Table \ref{table1}.} with the instrument response matrix. And $s$ is the intrinsic amplitude of the source. Under the background-only hypothesis (${\rm H}_{0}$), the probability is:
\begin{equation}
{{P}_{k}({d}_{k}|{H}_{0})}= \prod_{i} \frac{1}{\sqrt{2\pi}{\sigma}_{{n}_{i}}}exp(-\frac{{\widetilde{d}_{i}}^2}{2\sigma^{2}_{{n}_{i}}}).
\end{equation}
The log-likelihood ratio (LR) per detector is then computed as:
\begin{equation}
{ \mathcal L }_{k} = {\rm ln} \frac{{P}_{k}({d}_{k}|{H}_{1})}{{P}_{k}({d}_{k}|{H}_{0})} = \sum_{i=1}[{\rm ln}\frac{{\sigma}_{{n}_{i}}}{{\sigma}_{{d}_{i}}} + \frac{{\widetilde{d}_{i}}^2}{2\sigma^{2}_{{n}_{i}}} - \frac{({\widetilde{d}_{i}-{r}_{i}s)^2}}{2\sigma^{2}_{{d}_{i}}}],
\end{equation}
and the joint LR is: 
\begin{equation}
{ \mathcal L } = \sum_{k=1}{ \mathcal L }_{k}.
\end{equation}
The LR is computed for each detector and summed to obtain the joint LR, which serves as a test statistic quantifying the significance of the signal hypothesis. Given a known source location, the LR is evaluated for that specific direction. See also \citet{2025ApJS..277....9C} for the complete derivation and methodology.

To identify significant candidates, we adopt a detection threshold on the joint LR, empirically determined through extensive background-only simulations. Specifically, we performed $10^5$ targeted searches using synthetic light curves containing only Poisson fluctuation. In each simulation, we applied the same search pipeline with known source directions and recorded the joint LR to construct the background distribution \citep{2025ApJS..277....9C}.
This procedure also enables calculation of $p$-values for observed burst events, defined as the fraction of background simulations where the LR exceeds the observed value, providing a quantitative measure of the significance of the burst.

In addition, according to Wilks’s theorem, the LR statistic is expected to approximately follow a $\chi^2$ distribution under the null hypothesis \citep{2018ApJ...862..152K}. For a single degree of freedom, this provides an analytical estimate of the LR value corresponding to a given significance level. For example, a one-sided $3\sigma$ rejection threshold corresponds to an LR of approximately 9, which serves as a useful reference for assessing detection significance. Our previous studies \citep{2023MNRAS.518.2005C}
have shown that LR values derived from background simulations are roughly consistent with those predicted by Wilks’s theorem, particularly at high significance levels, supporting the robustness of this detection approach.

\citet{2025ApJS..277....9C} showed that the background distributions from joint and individual instrument searches are statistically consistent, in agreement with theoretical expectations. Based on this,
we ultimately adopt a relatively strict detection threshold based on empirical experience from the GECAM search pipeline, as reported in Table 1 of \citet{2025SCPMA..6839511C}. 

\subsection{Burst Identification}
Using the targeted search method described above, we identified a total of 116 candidate triggers potentially associated with GBM sub-threshold events. The distribution of the triggered joint LR values for these candidates is shown in Figure~\ref{fig10}. To obtain a more stringent and reliable sample, we performed a refined analysis for each candidate.

In the likelihood ratio framework, an unknown physical parameter, the source amplitude, is estimated by maximizing the likelihood function. For an event detected only by GBM, the amplitude optimized with GBM data alone (denoted as s$_{1}$) leads to a high LR value for GBM. If the same event is absent in GECAM, the amplitude estimated using GECAM data alone would be zero (s$_{2}=0$), resulting in an LR of zero for GECAM. When combining data from both instruments, the joint likelihood maximization balances the contributions from both detectors. In such cases, the amplitude derived from the joint analysis s$_{3}$ becomes smaller than s$_{1}$, educing the joint LR relative to the GBM-only LR. This effect, shown in Section 2.4 of the ETJASMIN pipeline paper \citep{2025ApJS..277....9C}, is typical for local events, such as particle events or background fluctuations, that are detected only by a single instrument.

To exclude such cases, we compared the GBM-only LR with the joint LR for each candidate. If the GBM LR exceeded the joint LR, the candidate was removed as a likely local or single-instrument event. For example, one event on 2024 December 9 at 05:35:44 UTC showed a high LR in GBM ($\sim$75) but no signal in GECAM ($\sim$0), resulting in a reduced joint LR of 37; such an event was excluded from our sample. Using this criterion, we removed 26 candidates.

In addition to the joint LR suppression effect described above, other factors may also lead to such discrepancies. Large localization uncertainties in GBM may place the true source position outside the field of view of the other satellite, preventing a coincident detection. Differences in detector sensitivity and the incident angle of the source may also result in detection by only one instrument. Non-astrophysical events, such as charged particle events (CP), can also produce spurious signals in a single instrument. Given the different orbital configurations of the satellites, CP events are rarely observed simultaneously by multiple spacecrafts.

To ensure the reliability of the remaining candidates, we manually examined cases where the GBM-only LR and the joint LR were close but potentially affected by noise or marginal signals. Certain cases with similar LR values may involve background fluctuations that cannot be fully distinguished by LR values alone. Therefore, we conducted a manual inspection guided by empirical judgment, using the light curves and sky maps to exclude false or ambiguous triggers. After this screening, we retained 49 events with coincident detections in more than one instrument for further analysis.

\begin{figure}[htbp]
\epsscale{1}
\plotone{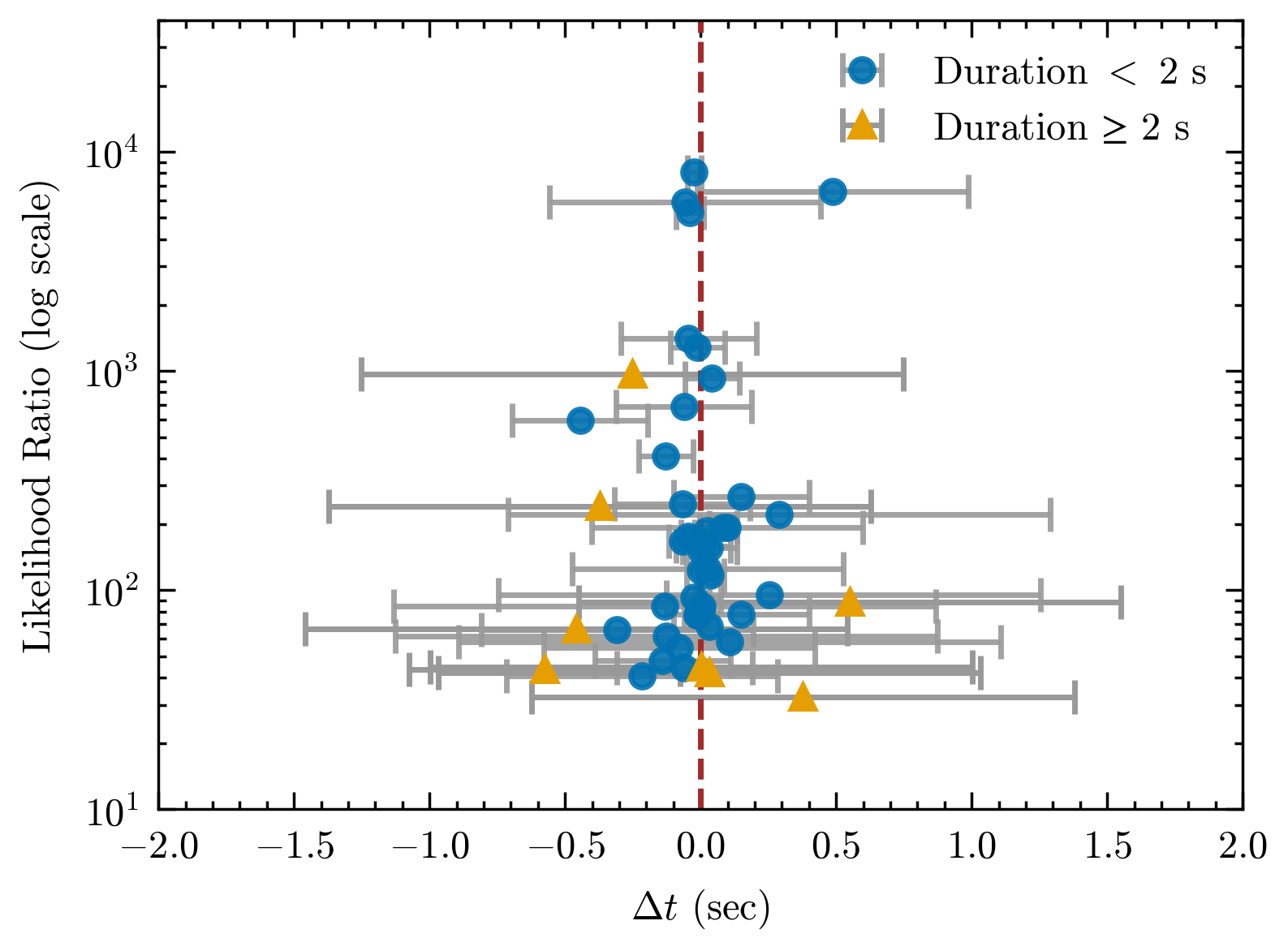}
\caption{Likelihood ratio vs. the time offset ($\Delta t$) between the GBM trigger time and the center of the detection window that resulted in the highest signal significance for the burst in our sample. Events are separated by duration: short-duration bursts ($<$2 s, blue circles) and long-duration bursts ($\geq$2 s, yellow triangles). Error bars represent the width of the detection window.}
\label{fig7}
\end{figure}

\section{Results}
\label{sec:results}

As described above, we identified 49 burst candidates from the GBM sub-threshold trigger catalogue over a period of about four years. These events passed the joint likelihood ratio threshold and were confirmed through multi-instrument coincidence checks and manual inspections to ensure their reliability.

Figure \ref{fig1} shows the distribution of joint likelihood ratios (LRs) for the 49 bursts with confirmed multi-instrument detections, grouped by GBM reliability score. We find that events with higher reliability (REL$= $8) tend to have substantially larger LR values, with the majority exceeding $10^2$. In contrast, lower-reliability events (REL$=$5) cluster at lower LR values, suggesting a positive correlation between the assigned reliability score and the multi-instrument detection significance.

Figure \ref{fig2} displays the light curves of a representative burst detected by all three instruments. A clear excess in count rate is observed near $T_0$ across GBM, GECAM-B, and GECAM-C, indicating a temporally coincident signal. Figures \ref{fig3} and \ref{fig4} present additional examples of sub-threshold events with confirmed multi-instrument detections. In both cases, the bursts exhibit temporally coincident excesses in at least two instruments, despite their relatively low signal strengths. 

Figure \ref{fig5} presents an example of a sub-threshold burst observed jointly by GBM, GECAM-B and GECAM-C on 2022 December 15. The left panels show the summed count rates across all detectors for each instrument, where no prominent excess is visible near $T_0$. In contrast, the right panels display the optimized weighted light curves, constructed by summing counts from all detectors and energy channels with appropriate weight factors (the ratio of expected source counts to background variance for each channel and detector), following the method described in Section 2.1 of \citealt{2016ApJ...826L...6C} and Section 4.3 of \citealt{2021MNRAS.508.3910C}. A clear, temporally coincident excess emerges in the optimized light curves, with the shaded region indicating the 
time bin that yields the highest signal significance. The joint significance of the signal is calculated based on the optimized weighted light curves \citep{2021MNRAS.508.3910C}.

Figure \ref{fig7} shows the distribution of likelihood ratio as a function of time offset ($\Delta$t) between the GBM trigger time and the center of the detection window with the highest LR. The sample includes both short-duration bursts (blue circles) and long-duration bursts (yellow triangles). Most events cluster near $\Delta t = 0$ s, as expected for signals temporally coincident with the GBM trigger.

The results of the targeted search for all 49 bursts are summarized in Table~\ref{table1}. The table lists, for each event: the event ID, GBM trigger number, date and time, burst duration, reliability score, time offset ($\Delta t$), search timescale, log-likelihood ratio ($\mathcal{L}$), signal-to-noise ratio (SNR), best-fit spectral template, and detection status in GECAM-B and GECAM-C.

\section{DISCUSSION}
\label{sec:discussion}

\subsection{Short Sub-threshold GRB Detection Rate}
Among the 466 GBM sub-threshold events with REL $\geq$ 5, 181 were identified as within GECAM’s field of view, corresponding to an overall visibility fraction of $\sim$39\%. This fraction varies by REL, with $\sim$17\% of REL = 5 and $\sim$22\% of REL = 8 events identified as visible. These estimates are based on the central localization positions reported by GBM and do not account for the statistical uncertainties of sub-threshold events. Accordingly, the true number of events within GECAM’s sky coverage is likely underestimated, and the reported visibility fraction is an approximate lower bound.

Joint likelihood analysis of these 181 events yielded 116 candidates with potential signals. However, manual inspection of light curves revealed several cases in which only one instrument detected a signal, leading to reduced joint detection significance. These cases may arise from the large localization uncertainties of GBM sub-threshold events, where the true source position lies outside GECAM’s actual field of view despite the central localization being nominally visible. In other cases, the signal may have arrived at an unfavorable angle or with insufficient intensity for detection. Some events may also originate from non-astrophysical sources such as charged particle events, which are typically localized and not detected simultaneously by multiple satellites due to differences in orbital environment and shielding. After excluding such ambiguous cases, we identified 49 events as confirmed burst candidates, indicating that at least $\sim$27\% (49/181) of the visible GBM sub-threshold events are consistent with real astrophysical transients jointly observed by GECAM.

\begin{figure}[htbp]
\epsscale{1}
\plotone{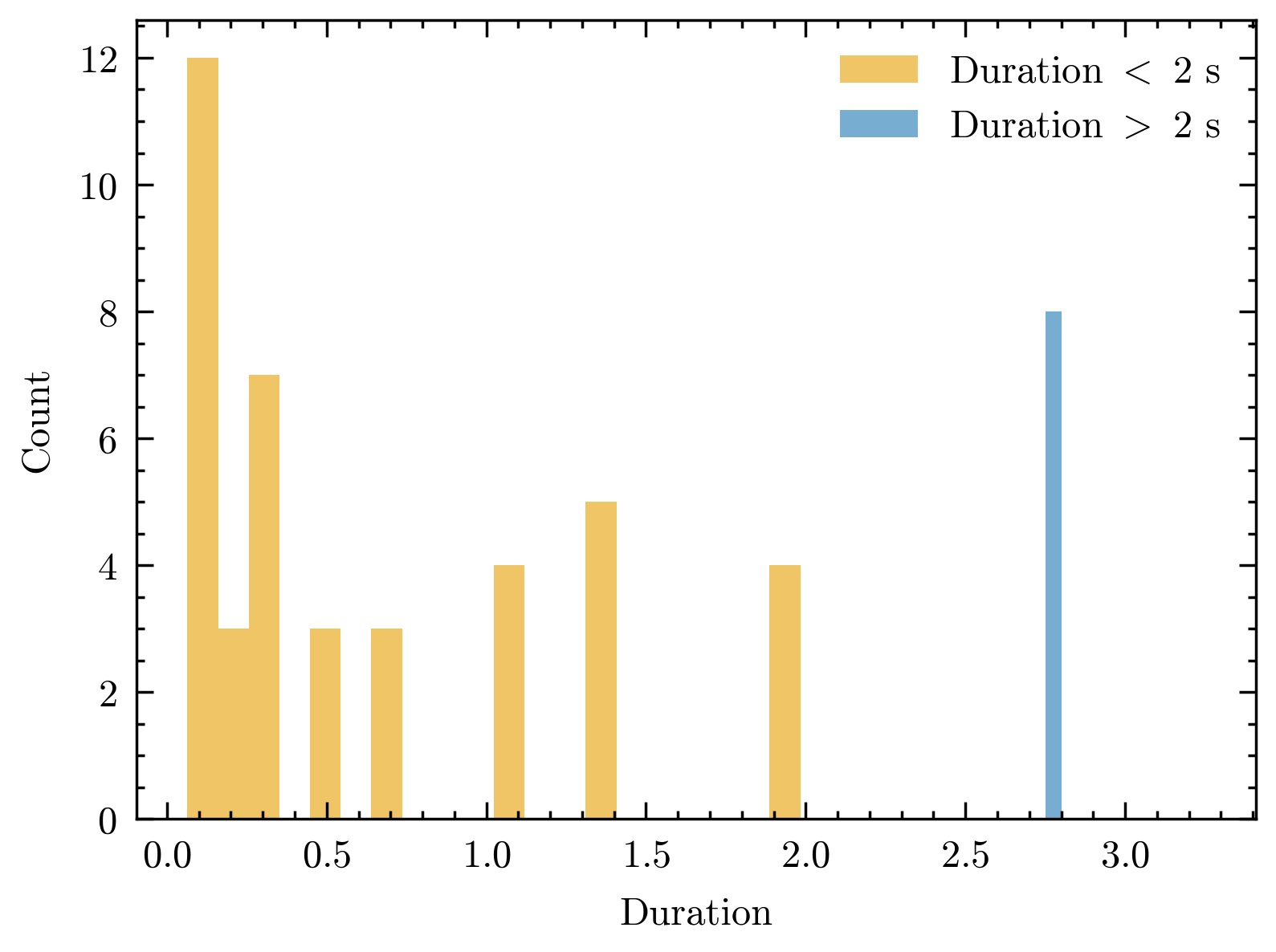}
\caption{Distribution of burst durations for the 49 sub-threshold events. Orange bars represent short-duration bursts ($<$2 s), and blue bars represent long-duration bursts ($\geq$2 s).}
\label{fig8}
\end{figure}

As shown in Figure \ref{fig1}, 39 of these 49 confirmed burst candidates have REL $=$ 8, corresponding to a confirmation rate of $\sim$38\% (39/102) among all REL $=$ 8 events considered. This indicates that a substantial fraction of high-REL candidates are real astrophysical bursts. Among these, 36 are classified as short GRBs, yielding a short-burst fraction of $\sim$35\% (36/102). For REL $=$ 5 events, 10 were identified in more than 1 instrument, implying a real astrophysical origin, corresponding to a confirmation rate of $\sim$13\% (10/79), with 5 classified as short GRBs (short-burst fraction $\sim$6\%). These results show that REL = 8 events represent the major contributor to the confirmed sample, reflecting their higher reliability and higher short-burst occurrence.

We note that GBM triggers on $\sim$240 GRBs annually, including $\sim$40 short GRBs \citep{2020ApJ...893...46V}. In our sample, we identified 41 short-duration GRBs among the confirmed joint detections (Figure \ref{fig8}), corresponding to an average of $\sim$10 additional GECAM-confirmed short GRBs per year over the four-year span. Importantly, this estimate reflects only those events that are both visible to and detectable by GECAM. Given GECAM’s limited sky coverage and detection sensitivity, a number of short GRBs within the sub-threshold population likely remain undetected. 

However, we could extrapolated the observed confirmation rates to the full catalog. During the 8 years of GBM subthreshold triggers, among the 709 REL = 8 events, we estimate $\sim$248 short GRBs (709 × 0.35), and among the 1348 REL = 5 events, $\sim$80 short GRBs (1348 × 0.06), yielding a total of $\sim$41
short GRBs per year. 
This suggests that a future complete and systematic joint monitoring of GBM sub-threshold events can enhance the GBM annual detection rate of short GRBs by $\sim$100\% relative to the GBM-triggered sample. These results highlight the importance of systematic, multi-instrument searches in expanding the observable short GRB population and enhancing the completeness of current GRB catalogs. This is particularly relevant for short-duration bursts, which play an important role in studies of compact object mergers and multi-messenger astrophysics.

\subsection{Properties of Confirmed Burst Candidates}

Representative light curves of selected events (Figures \ref{fig2}–\ref{fig5}) demonstrate that a subset of sub-threshold events exhibit clear, temporally coincident excesses across multiple instruments, even in cases where the overall signal strength is low. While some bursts show visible excesses in summed detector counts (e.g., Figure \ref{fig2}), others reveal significant signals only after applying optimized weighting across all detectors (e.g., Figure \ref{fig5})). These findings highlight the effectiveness of multi-detector analysis in recovering weak events that may not trigger onboard algorithms.

The temporal distribution of detection windows (Figure \ref{fig7}) shows that most events are centered near the GBM trigger time, consistent with coincident burst signatures. Some events show time offsets from the GBM trigger, which may result from differences between multi-instruments in detection response.

\subsection{Search for SGRB–GW Association}
Gravitational-wave detections have enabled multi-messenger studies, where identifying high-energy electromagnetic counterparts is essential for probing compact object mergers. In the first three observing runs (O1 to O3; \citealt{2019PhRvX...9c1040A,2021PhRvX..11b1053A, PhysRevX.13.041039}), only one such counterpart was confirmed: GW170817, a binary neutron star merger detected by gravitational-wave interferometers (e.g., LIGO \cite{2015CQGra..32g4001L}, Virgo \cite{2015CQGra..32b4001A} and KAGRA \cite{2021PTEP.2021eA101A}) and by gamma-ray instruments \citep{2017PhRvL.119p1101A, 2017ApJ...848L..14G, 2017ApJ...848L..15S}. The fourth observing run (O4), which began in May 2023 and is expected to continue through November 2025\footnote{\url{https://observing.docs.ligo.org/plan/index.html}}, has so far produced approximately 200 candidate GW events as reported by the LIGO--Virgo--KAGRA collaboration \footnote{\url{https://gracedb.ligo.org/superevents/public/O4/}}.

To explore potential associations between short gamma-ray bursts and gravitational-wave events, we examined temporal matches between 13 short-duration GRB candidates from our confirmed sample, identified since the start of O4, and publicly reported significant GW candidates\footnote{\url{https://gracedb.ligo.org/superevents/public/O4/}}. No coincidences were found within standard time windows ($-30$ to $+30$ seconds relative to the GRB trigger time), as confirmed by similar searches \citep{2020ApJ...893..100H,2024ApJ...964..149F}. While no temporal matches were identified, the result highlights the importance of coordinated multi-messenger strategies in future observing runs. Improved pipelines, such as ETJASMIN, are expected to increase the likelihood of detecting faint or sub-threshold gamma-ray counterparts and thereby strengthen the prospects for joint detections.

\section{Summary}
\label{sec:summary}
In this work, we present a systematic search and verification of these SGRB candidates from the Fermi/GBM sub-threshold triggers by jointly analyzing data from GECAM-B and GECAM-C. Among 466 Fermi/GBM sub-threshold events (with reliability $\geq$5) from 2021 to 2024, 181 are within GECAM’s field of view. We find that 49 out of 181 are 
confirmed astrophysical transients. These events exhibit coherent excesses across multiple detectors and span a broad range of likelihood ratios. 
And 41 of 49 ($\sim$84\%) events can be classified as SGRBs.
 
With these GECAM-recovered SGRB events, the SGRB detection rate of Fermi/GBM is increased from about 40 per year (only triggered events) to about 50 per year (GBM triggered and GECAM-recovered events from GBM sub-threshold triggers). These results suggest that a future multi-instrument complete monitoring and systematic verification of GBM sub-threshold triggers is expected to increase the detectable SGRB rate to about 80 per year, by $\sim$100\% improvement
relative to GBM-triggered events. This will provide important improvement to the current SGRB catalogs and improving the completeness of the observed GRB population. Moreover, we note that this improved SGRB detection rate may have important implications on the estimation of the local formation rate of SGRB and the binary merger rate. However, in-depth studies are needed on these topics.

Lastly, we also searched for the temporal coincidence between SGRBs and gravitational wave (GW) from the LIGO–Virgo–KAGRA O4 run. However, we did not find any coincident case, indicating the rarity of the multi-messenger event. 
These results highlight the necessity of coordinated observation and multi-instrument analysis pipelines, such as ETJASMIN, in improving the sensitivity to faint gamma-ray transients with the current instruments.

\begin{longrotatetable}
\centering
\begin{deluxetable*}{lccccccccccccc}
\tablecaption{Properties of the Jointly Confirmed Sub-threshold SGRB Candidates from Fermi/GBM, GECAM-B and GECAM-C\label{table1}}
\tablewidth{700pt}
\tabletypesize{\scriptsize}
\tablehead{
\colhead{ID} & \colhead{TrigNum $^{1}$} & 
\colhead{Date $^{1}$} & \colhead{Time$^{1}$} & 
\colhead{Duration$^{1}$ (s)} &  \colhead{Rel$^{1}$} & \colhead{$\Delta t$ $^2$ (s)} &  \colhead{Timescale (s)} &
\colhead{$\mathcal L$} & \colhead{SNR} &
\colhead{Template} &  \colhead{GECAM-B $^3$} & \colhead{GECAM-C $^3$}}
\startdata
1	&	639199407	&	2021-04-04	&	3:23:22.40	&	0.32	&	8	&	0.009 	&	0.2	&	158.72	&	17.44	&	RspHard	&	1	&	0	\\
2	&	651339138	&	2021-08-22	&	15:32:13.90	&	0.128	&	5	&	-0.013 	&	0.05	&	77.19	&	12.32	&	RspSoft	&	1	&	0	\\
3	&	653118262	&	2021-09-12	&	5:44:17.08	&	0.32	&	8	&	0.036 	&	0.1	&	116.7	&	15.22	&	RspSoft	&	1	&	0	\\
4	&	655868466	&	2021-10-14	&	1:41:01.96	&	0.32	&	8	&	-0.059 	&	0.5	&	44.22	&	9.36	&	RspNorm	&	1	&	0	\\
5	&	656732056	&	2021-10-24	&	1:34:11.42	&	0.096	&	8	&	0.034 	&	0.2	&	155.96	&	17.53	&	RspNorm	&	1	&	0	\\
6	&	662385068	&	2021-12-28	&	11:51:03.00	&	2.751	&	5	&	0.003 	&	2	&	44.65	&	9.49	&	RspSoft	&	1	&	0	\\
7	&	662497751	&	2021-12-29	&	19:09:06.44	&	0.064	&	5	&	-0.003 	&	0.05	&	166.1	&	17.81	&	RspSoft	&	1	&	0	\\
8	&	663645963	&	2022-01-12	&	2:05:58.31	&	0.192	&	8	&	0.083 	&	0.1	&	193.08	&	19.55	&	RspSoft	&	1	&	0	\\
9	&	663703033	&	2022-01-12	&	17:57:08.26	&	0.512	&	8	&	-0.128 	&	0.2	&	409.27	&	28.38	&	RspSoft	&	1	&	0	\\
10	&	663784002	&	2022-01-13	&	16:26:37.56	&	0.064	&	8	&	-0.002 	&	0.1	&	123.32	&	15.56	&	RspSoft	&	1	&	0	\\
11	&	663926088	&	2022-01-15	&	7:54:43.88	&	0.32	&	5	&	-0.067 	&	0.1	&	167.91	&	18.09	&	RspSoft	&	1	&	0	\\
12	&	665237697	&	2022-01-30	&	12:14:52.36	&	2.751	&	5	&	0.378 	&	2	&	32.62	&	8.17	&	RspNorm	&	1	&	0	\\
13	&	669921389	&	2022-03-25	&	17:16:24.55	&	2.751	&	5	&	0.551 	&	2	&	88.22	&	13.25	&	RspSoft	&	1	&	0	\\
14	&	671515269	&	2022-04-13	&	4:01:04.94	&	1.024	&	8	&	-0.077 	&	1	&	54.54	&	10.42	&	RspHard	&	1	&	0	\\
15	&	671989925	&	2022-04-18	&	15:52:00.20	&	0.32	&	8	&	-0.061 	&	0.5	&	686.36	&	36.73	&	RspHard	&	1	&	0	\\
16	&	674879945	&	2022-05-22	&	2:39:00.75	&	2.751	&	8	&	-0.251 	&	2	&	968.01	&	43.66	&	RspNorm	&	1	&	0	\\
17	&	675265117	&	2022-05-26	&	13:38:32.58	&	0.128	&	8	&	-0.046 	&	0.05	&	175.93	&	18.45	&	RspSoft	&	1	&	0	\\
18	&	675411172	&	2022-05-28	&	6:12:47.69	&	0.703	&	8	&	-0.045 	&	0.5	&	1406.91	&	51.19	&	RspHard	&	1	&	0	\\
19	&	677691241	&	2022-06-23	&	15:33:56.56	&	1.407	&	8	&	-0.444 	&	0.5	&	595.24	&	34.4	&	RspHard	&	1	&	0	\\
20	&	677748900	&	2022-06-24	&	7:34:55.66	&	1.024	&	8	&	0.149 	&	0.5	&	77.68	&	12.38	&	RspNorm	&	1	&	0	\\
21	&	679214831	&	2022-07-11	&	6:47:06.28	&	1.983	&	8	&	0.289 	&	2	&	221.89	&	20.91	&	RspNorm	&	1	&	0	\\
22	&	679248723	&	2022-07-11	&	16:11:58.60	&	0.192	&	8	&	-0.011 	&	0.2	&	1277.82	&	49.95	&	RspHard	&	1	&	0	\\
23	&	679971925	&	2022-07-20	&	1:05:20.63	&	2.751	&	8	&	-0.372 	&	2	&	240.95	&	21.87	&	RspSoft	&	1	&	0	\\
24	&	680056679	&	2022-07-21	&	0:37:54.60	&	1.407	&	8	&	0.098 	&	1	&	193.75	&	19.63	&	RspHard	&	1	&	0	\\
25	&	680782656	&	2022-07-29	&	10:17:31.69	&	0.512	&	8	&	-0.068 	&	0.5	&	248.55	&	22.07	&	RspNorm	&	1	&	0	\\
26	&	682732651	&	2022-08-20	&	23:57:26.20	&	0.064	&	8	&	0.007 	&	0.05	&	84.3	&	12.91	&	RspNorm	&	1	&	0	\\
27	&	692785392	&	2022-12-15	&	8:23:07.35	&	1.983	&	8	&	-0.140 	&	0.5	&	47.96	&	9.63	&	RspHard	&	1	&	1	\\
28	&	694750491	&	2023-01-07	&	2:14:46.70	&	0.703	&	8	&	-0.308 	&	1	&	66.24	&	11.63	&	RspNorm	&	0	&	1	\\
29	&	697172795	&	2023-02-04	&	3:06:30.42	&	0.096	&	8	&	0.003 	&	0.05	&	150.67	&	16.73	&	RspNorm	&	0	&	1	\\
30	&	698570141	&	2023-02-20	&	7:15:36.30	&	1.407	&	5	&	-0.215 	&	1	&	40.62	&	8.94	&	RspSoft	&	0	&	1	\\
31	&	698655879	&	2023-02-21	&	7:04:34.54	&	2.751	&	8	&	-0.459 	&	2	&	66.9	&	11.49	&	RspSoft	&	0	&	1	\\
32	&	698668405	&	2023-02-21	&	10:33:20.04	&	2.751	&	5	&	0.033 	&	2	&	42.03	&	9.22	&	RspSoft	&	0	&	1	\\
33	&	698759889	&	2023-02-22	&	11:58:04.87	&	1.407	&	5	&	-0.133 	&	2	&	84.53	&	12.94	&	RspSoft	&	0	&	1	\\
34	&	698760101	&	2023-02-22	&	12:01:36.26	&	1.983	&	8	&	0.255 	&	2	&	95.3	&	13.67	&	RspSoft	&	0	&	1	\\
35	&	699106080	&	2023-02-26	&	12:07:55.43	&	2.751	&	5	&	-0.576 	&	1	&	43.36	&	9.32	&	RspSoft	&	0	&	1	\\
36	&	699811746	&	2023-03-06	&	16:09:01.45	&	0.128	&	8	&	0.034 	&	0.05	&	68.11	&	11.62	&	RspNorm	&	1	&	1	\\
37	&	720550737	&	2023-11-01	&	16:58:52.25	&	0.096	&	8	&	-0.011 	&	0.1	&	77.18	&	12.41	&	RspNorm	&	1	&	0	\\
38	&	721431316	&	2023-11-11	&	21:35:11.38	&	0.064	&	8	&	0.022 	&	0.05	&	186.61	&	18.68	&	RspSoft	&	1	&	0	\\
39	&	726465700	&	2024-01-09	&	4:01:35.16	&	0.32	&	8	&	-0.025 	&	0.2	&	93.04	&	13.04	&	RspNorm	&	1	&	1	\\
40	&	728111754	&	2024-01-28	&	5:15:49.88	&	1.983	&	8	&	-0.128 	&	2	&	61.71	&	11.16	&	RspSoft	&	1	&	0	\\
41	&	729327521	&	2024-02-11	&	6:58:36.00	&	1.407	&	8	&	0.108 	&	2	&	58.29	&	10.77	&	RspNorm	&	0	&	1	\\
42	&	740166710	&	2024-06-15	&	17:51:45.00	&	0.064	&	8	&	-0.023 	&	0.05	&	8093.29	&	120.06	&	RspHard	&	0	&	1	\\
43	&	742324312	&	2024-07-10	&	17:11:47.50	&	0.703	&	8	&	0.025 	&	1	&	125.33	&	15.66	&	RspSoft	&	0	&	1	\\
44	&	742715048	&	2024-07-15	&	5:44:03.30	&	0.192	&	8	&	-0.039 	&	0.1	&	5264.27	&	96.73	&	RspHard	&	0	&	1	\\
45	&	745016845	&	2024-08-10	&	21:07:20.84	&	0.32	&	8	&	0.042 	&	0.2	&	926.82	&	41.62	&	RspNorm	&	0	&	1	\\
46	&	745924031	&	2024-08-21	&	9:07:06.78	&	0.096	&	8	&	0.002 	&	0.05	&	80.89	&	12.85	&	RspSoft	&	0	&	1	\\
47	&	745958167	&	2024-08-21	&	18:36:02.93	&	1.024	&	8	&	-0.057 	&	1	&	5891.26	&	99.4	&	RspNorm	&	0	&	1	\\
48	&	753774845	&	2024-11-20	&	5:54:01.00	&	1.024	&	8	&	0.486 	&	1	&	6600.44	&	109.05	&	RspNorm	&	0	&	1	\\
49	&	755576606	&	2024-12-11	&	2:23:21.90	&	0.512	&	8	&	0.150 	&	0.5	&	266.23	&	22.97	&	RspNorm	&	1	&	0	\\
\enddata
\tablecomments{$^1$ GBM trigger information from the sub-threshold archive. \\
$^2$ Time offset between the GBM trigger and the center of the GECAM detection window with the highest significance.  \\
$^3$ joint detection flag (1 = within field of view and detected).}
\end{deluxetable*}
\end{longrotatetable}

\section*{Acknowledgments}
We thank the anonymous reviewer for very helpful comments and suggestions. This work is supported by the National Natural Science Foundation of China 
(Grant Nos. 12303045, 
12273042, 12494572, 
12373047) 
, the National Key R\&D Program of China
(2022YFF0711404, 
2021YFA0718500
), 
the Strategic Priority Research Program of Chinese Academy of Sciences (Grant Nos. 
XDA30050000, 
XDB0550300
). 
The GECAM (Huairou-1) mission is supported by the Strategic Priority Research Program on Space Science (Grant No. XDA15360000) of the Chinese Academy of Sciences.
C.C. acknowledges the support from Hebei Natural Science Foundation (No. A2023205020) and the Science Foundation of Hebei Normal University (No. L2023B11). 
We appreciate the public data of Fermi/GBM. We thank the development and operation teams of GECAM.

\bibliography{ref}
\bibliographystyle{aasjournal}

\end{document}